\documentclass[aps,pra,preprint,showpacs]{revtex4}
\usepackage{amsmath,graphicx}
\begin{document}
\def\la{{\langle}}
\def\ra{{\rangle}}
\def\q{{\quad}}
\def\[{{\{}}
\def\]{{\}}}

\title{Path integrals, the ABL rule and the three-box paradox}
%
%
\author{D.Sokolovski*, I.Puerto Gim\'enez$^\dagger$ and R. Sala Mayato$^\dagger$$^\ddagger$}

\address{$\it *$ School of Mathematics and Physics,
  Queen's University of Belfast,
  Belfast, BT7 1NN, United Kingdom,}

\address{$^\dagger$ Departamento de F\'\i sica Fundamental II,
   Universidad de La Laguna,
   La Laguna (S/C de Tenerife), Spain}

\address{$^\ddagger$ IUdEA (Instituto Universitario de Estudios Avanzados)
   Universidad de La Laguna,
   La Laguna (S/C de Tenerife), Spain}

{\it * Corresponding author, e-mail d.sokolovski@qub.ac.uk, ph.:  +44 28 90 976055,
fax: +44 28 90 976061}
 
\date{\today}

\begin{abstract}
The three-box problem is analysed 
 in  terms of virtual pathways, interference
 between which is destroyed by 
 a number of  intermediate
 measurements. The Aharonov-Bergmann-Lebowitz (ABL)
rule is shown to be a particular case of Feynman's
recipe for assigning probabilities to exclusive
alternatives.
The 'paradoxical' features of the three box case
arise in an attempt to attribute, in contradiction to the 
uncertainty principle, properties 
pertaining to different ensembles
produced by different intermediate measurements to the 
same particle. The effect can be 
mimicked by a classical system, provided an observation
is made to perturb the system in a non-local manner.

\end{abstract}

%
%
\pacs{PACS number(s): 03.65.Ta, 73.40.Gk}

\maketitle

\section{Introduction}
The three-box paradox first introduced by Aharonov
and Vaidman \cite{3B1} 
concerns results of intermediate measurements
performed on the pre- and post-selected three-level quantum system.
The authors of \cite{3B1,ABOOK,3B2}, who considered the problem within the 
two-state vector formalism of quantum mechanics (see, for example, Ref.\cite{ABOOK}),
 noted 
that a quantum particle can, in some sense, exist with certainty in two different
boxes  at the same time.
They also suggested  that quantally the product rule (the product of two sharply
defined variables $a$ and $b$  is a sharply defined variable
with the value $ab$) may be violated.
As the main tool for calculating probabilities they used the
Aharonov-Bergmann-Lebowitz (ABL) formula \cite{ABL1}
extended in \cite{3B1} to include variables with degenerate 
eigenvalues.
\newline 
There continues to be a debate regarding paradoxical 
(or otherwise) status of the three-box case \cite{Leif}-\cite{Fink}.
In Ref.\cite{Kast}  it was pointed out
that the paradox arises from combining properties
which cannot be possessed by the same individual particle.
Refs. \cite{Kirk2,Kirk3} note the analogy with
a three-slit diffraction experiment and demonstrate
the existence of a purely  classical stochastic model 
with similar properties.  The authors
of Ref.\cite{Leav} find, 
 in somewhat similar vein, the 'non-paradoxical' roots
 of the three-box phenomenon in the destructive interference.
 \newline
An analysis of what appears  to be a quantum 'paradox' 
usually benefits from choosing a simple yet general language.
In this paper we propose Feynman's formulation of  quantum mechanics 
as an alternative to the approach of Refs. \cite{3B1,ABOOK,3B2} .
Feynman's analysis focuses on the double slit experiment which captures
the essence of all quantum interference phenomena.
The double slit experiment cannot itself be "explained',
yet many phenomena in quantum mechanics can be reduced to it
directly or indirectly. 
A simple yet complete set of axioms apply to any system
capable of reaching final state(s) via a number of alternative routes:
\newline
(I) ({\it Uncertainty 
Principle}) \cite{Feyn1}: ' Any determination of the alternative taken by a process
capable of following more that one alternative
destroys the interference between alternatives", from which
it follows that interfering alternatives cannot be told apart and form, therefore, a single
indivisible pathway.
\newline
 (II) ({\it Recipe for Assigning Probabilities})  (Ref. \cite{Feyn2},
p.1-10): "When an event can occur in several alternative ways, the probability
amplitude for the event is the sum of the probability amplitudes for each way  considered
separately. If an experiment is performed which is capable 
of determining when one or another alternative is actually taken,
the probability of the event is the sum of the probabilities of each alternative'.
\newline
Feynman path analysis has been applied to the Hardy's paradox in \cite{SPS}
and in this paper we will use it to analyse measurements
performed on pre- and post-selected systems and in particular to the three box paradox mentioned
above.
The rest of the paper is organised as follows.
In Section 2 we follow Refs. \cite{SR1,SR2} in considering a
measurement as an act of destruction of interference between
certain virtual pathways.
In Sect.3 we show the the ABL rule to be a particular consequence
of the Feynman's recipe for assigning probabilities.
In Sect.4 we analyse the original
three-box case.
Section 5 discusses a classical model designed to mimic the quantum three-box case.
Section 6  contains our conclusions.

\section {Measurements, virtual paths and classical stochastic networks}
To apply the Feynman rule (II) one must specify the alternatives, the corresponding
probability amplitudes and the means by which interference between the 
alternatives  can be destroyed. 
In a textbook attempt  to determine the slit chosen by an electron in a double slit experiment, 
one must destroy coherence between 
the paths passing through different slits, e.g., by illuminating it with photons \cite{Feyn2}.
To apply a similar reasoning to
We start by identifying the alternatives which can be de-cohered
by an additional
intermediate measurement performed on a quantum system pre- and post-selected in 
given initial and final states.
 A general answer to this question has been given in Refs.\cite{SR1,SR2}.
For a quantum system in an $N$-dimensional Hilbert
space with an orthonormal basis $\{|n\ra\}, \quad n=1,2..N$,
a transition amplitude between arbitrary initial and final states $|i\ra$ and $|f\ra$ can be written  as a sum 
over {\it virtual} (i.e., interfering) paths \cite{SR1}
\begin{equation} \label{0.2}
\la f|\exp(-i\hat{H} T)|i\ra = lim_{K\rightarrow \infty}\sum_{n_1,n_2,...n_K}\la f|\prod_{j=1}^K |n_j\ra
\la n_j|\exp(-i\hat{H} T/K)|i\ra \equiv \sum_{\[n\]}
\Phi^{f\leftarrow i}_{ \[n\]}.
\end{equation}
In equation (\ref{0.2}) the (Feynman) probability amplitude $\Phi^{f\leftarrow i}_{ \[n\]}$,
defined by the product sandwiched between the initial and
final states,  is summed over all irregular paths $n(t)$ which
at any given time may take values of $1,2,...N$.
 The best known example of a path sum similar to Eq.(\ref{0.2}) is the Feynman path integral \cite{Feyn1} over the paths defined in the 
 co-ordinate space of a point particle.
\newline
Assume next that between $t=0$ and $t=T$ the system is coupled to a pointer with position
$x$ prepared in an initial state $|M\ra$,  $\la x|M\ra= G(x)$ so that the total Hamiltonian is
\begin{equation} \label{0.4}
\hat{\mathcal{H}}= \hat{H}-i\partial_x \beta(t)F(\hat{n}),
\end{equation}
where $\beta(t)$ is some switching function and
\begin{equation} \label{0.3}
F(\hat{n})\equiv\sum_{n=1}^N |n\ra F(n) \la n|.
\end{equation}
is an operator, whose eigenvalues may or may not be degenerate, depending
on whether or not the function  $F(n)$ may take the same values for different values
of $n$. The probability amplitude for the system to be found in the final state $|f\ra$
and the pointer to have a final position $x$ is given by a restricted path sum \cite{SR1}
\begin{equation} \label{0.2a}
\la x|\la f|\exp(-i\hat{\mathcal{H}} T)|i\ra|M\ra = \sum_{\[n\]}G (x-\int_0^T
\beta(t)F(n)dt)
\Phi^{f\leftarrow i}_{ \[n\]}.
\end{equation}
Setting the initial pointer position to $0$ by choosing, for example, $G(x)\rightarrow 
(\alpha/\pi)^{1/4}\exp(-\alpha x^2)$, $|G(x)|^2 \rightarrow \delta(x)$, 
we obtain a generalisation of the conventional von Neumann measurement,
in which the pointer position correlates with the value of the functional
$F[n(t)]=\int_0^T\beta(t)F(n(t))dt$, defined on the Feynman paths of the system.
Different choices of $\beta(t)$ correspond to different choices of the measured quantity.
We will be interested in the simple case when an intermediate impulsive von Neumann measurement,
of an operator $F(\hat{n})$ with $M<N$ distinct eigenvalues $F_m$, $m=1,2...M$ is conducted on a system with a zero Hamiltonian, $\hat{H}=0$ at, say, $t=T/2$,
\begin{equation} \label{0.7}
\beta (t) = \delta (t-T/2).
\end{equation}
 At $t=0$ the system prepared in an initial state 
$|i\ra$ and at  $t=T$ it is registered in one of the states belonging
to an orthonormal basis $\{ |f\ra, |g\ra, |h\ra,...\}$. Now in Eq.(\ref{0.2}) there remain only $N$ constant Feynman paths,
$n(t)=1,2,...,N$ with the corresponding amplitudes given by 
\begin{equation} \label{0.8}
\Phi^{z\leftarrow i}_n=\la f|n\ra \la n|i\ra, \quad z=f,g,h,...
\end{equation}
and the measured quantity is just the value,
of $F(n(t))$ at $t=T/2$.
Without a meter the probability to find the system in a state $|z\ra$ at $t=T$ is given by
\begin{equation} \label{0.8a}
\tilde{w}^{z\leftarrow i} = |\sum_{n=1}^N \Phi^{z\leftarrow i}_n|^2, \quad \sum_z \tilde{w}^{z\leftarrow i}=1.
\end{equation}
To find the same probability for a system coupled to a meter we trace the pointer
variable out in a pure state corresponding to the amplitudes in Eq.(\ref{0.2a}) to obtain
\begin{equation} \label{0.8a}
{w}^{z\leftarrow i} = \sum_{m=1}^M|\sum_{n=1}^N\Delta(F_m-F(n)) \Phi^{z\leftarrow i}_n|^2, \quad \sum_z {w}^{z\leftarrow i}=1,
\end{equation}
where $\Delta(z)\equiv 1$ for $z=0$ and $0$ otherwise.

The measurement can now be described in the language of 
interfering alternatives. 
With no measurements made, $N$ Feynman paths connecting $|i\ra$ 
with the same final state $|z\ra$
 are interfering alternatives which,
in accordance with the uncertainty principle (I), should be treated as a single indivisible pathway.
Interaction with a meter produces $M$ non-interfering (i.e., {\it real}) pathways represented by the 
classes of constant virtual paths sharing the same value $F(n)$.
Coherence between the classes is destroyed since the paths with different values of  $F(n)$
lead to distinguishable outcomes - different pointer positions.
The probability amplitude for each class is just the sum of the Feynman's amplitudes for each individual path. With or without a meter, Feynman paths leading to distinguishable orthogonal final
states are exclusive alternatives.

Thus, together with post-selection, intermediate measurements produce
a classical stochastic network of non-interfering  real pathways which are travelled
with observable frequencies. 
However, unlike their classical counterparts,  quantum measurements are, in general, invasive
in the sense that a mere act of measurement {\it fabricates} a new network for each choice of the
measured quantity and in Section 4 we will use the three-box example to illustrate this point.
For a classical analogy of such a behaviour  one can envisage a purely classical system whose statistical
properties change depending of whether or not a particular observation observation has been made.
While classically we must must make special provisions in to make an observation
alter the system's evolution, in the quantum case such an alteration is provided, whether we want it or not, by the invasive action of the meter. A classical analogue of the three-box case will 
be discussed in Sect.5 and next we will show the ABL rule to be a particular case of Feynman's
recipe for assigning probabilities.

\section {Feynman's  recipe vs. the ABL rule}
Thus, an intermediate
measurement of an operator $F(\hat{n})$ with $M<N$ distinct eigenvalues $F_m$ has been shown to produce a network of $M$ routes leading to $N$ final states $|f\ra, |g\ra, |h\ra,...$. A probability can  be assigned to each route using the Feynman's recipe (II), i.e.,
by summing as appropriate the amplitudes (\ref{0.8}) and squaring the modulus
 of the result, which yields
\begin{equation} \label{3.1}
w^{z\leftarrow i}_m=|\sum_{n=1}^N\Delta(F_m-F(n))\Phi^{z\leftarrow i}_{ n}|^2, \quad
z=f,g,h....
\end{equation}
for the probability $w^{f\leftarrow i}_m$ to end up in a state
$|f\ra$ after obtaining a value $F_m$ in the intermediate measurement.
We may ask further, what is the probability $P^{f\leftarrow i}_m$ to obtain the value $F_m$ provided the system
is later observed in the state $|f\ra$? To answer this question the authors of Ref. \cite{3B1}, \cite{ABOOK} employed the two-state vector formalism of quantum mechanics. We, on the other hand, 
can obtain $P^{f\leftarrow i}_m$ simply by repeating the experiment many times, $N_{ex}$, and dividing the number of arrivals in
$|f\ra$ conditioned on obtaining the result $F_m$, $w^{z\leftarrow i}_m N_{ex}$ by the total number of arrivals in $|f\ra$ (\ref{0.8a}) , $w^{z\leftarrow i} N_{ex}$,
\begin{equation} \label{3.2}
P^{f\leftarrow i}_m=\frac{|\sum_{n=1}^N\Delta(F_m-F(n))\Phi^{f\leftarrow i}_{ n}|^2}
{\sum_{m=1}^M |\sum_{n=1}^N\Delta(F_m-F(n))\Phi^{f\leftarrow i}_{ n}|^2}.
\end{equation}
Taking into account Eq.(\ref{0.8}) and introducing projectors $\hat{P}_m \equiv
\sum_{n=1}^N |n\ra \Delta(F_m-F(n))\la n|$ onto the subspaces
of eigenstates corresponding to the same $F_m$ we can rewrite (\ref{3.2}) as
\begin{equation} \label{3.3}
P^{f\leftarrow i}_m=\frac{|\la f|\hat{P}_m|i\ra |^2}
{\sum_{m=1}^M |\la f|\hat{P}_m|i\ra |^2}
\end{equation}
which is the ABL formula, introduced in  \cite{ABL1} and extensively discussed in \cite{3B1,ABOOK} in the general context of the probability
theory. It can, therefore, be seen as a simple consequence of the Feynman's rule (II) applied to the 
to the entire classical network defined by the basis in which the post-selection is made as well as the set of intermediate measurements performed.

We continue with an example which, although of no direct importance for the discussion 
of the three-box case in the following Section,
serves to further illustrate the general nature of the Feynman's rule. 
Consider a two-level ($N=2$) system pre- and post-selected
in arbitrary states $|i\ra$ and $|f\ra$ at $t=0$ and $t=T$,
respectively. The hamiltonian of the system is zero except
for $ T/2 - \tau / 2 < t < T/2 + \tau / 2 $ when an interaction $ \hat{H} $ is switched on.
The interaction causes
 transitions between the states $|1\ra$ and $|2\ra$ and the corresponding
transitions amplitudes are
\begin{eqnarray} \label{3.4}
S_{jj'} = \la j|\exp(-i\hat{H}\tau)|j'\ra \quad j,j'=1,2.
\end{eqnarray}
Next we consider the dichotomic variable $\hat{n} =\sum_{n=1}^2 |n\ra n \la n|$
and set out to measure the difference $\Delta n$ between its values before and after the interaction
has acted, but in such a way that the values themselves remain indeterminate.
To do so we employ a single meter of the type discussed in the Sect.2 with the coupling
\begin{eqnarray} \label{3.5}
-i\partial_f[\delta(t-T/2-\tau/2)-\delta(t-T/2+\tau/2)]\hat{n}.
\end{eqnarray} 
The meter acts the first time just before the perturbation, and the pointer is
shifted by either $-1$ or $-2$. The same pointer is shifted again just after the 
perturbation by either $1$ or $2$ and ends up with a shift of either $0$, $1$ or $-1$,
indicating the three possible values of $\Delta n$.
It is difficult to see how an ABL formula similar to (\ref{3.3}), which employs 
projectors onto orthogonal sub-spaces associated with each 
measurement outcome \cite{3B1,ABOOK}, can describe
the probability of the outcomes now that their number (three) exceeds that  of
available sub-spaces (two). The approach based on Feynman's paths is,
on the other hand, straightforward.
There are four virtual paths connecting $|i\ra$ and $|f\ra$. Two
correspond to the system remaining in the same state all along, 
have the value $\Delta n=0$ and cannot be distinguished by our measurement.
The corresponding probability amplitudes are
\begin{eqnarray} \label{3.6}
\Phi^{f\leftarrow i}_{11}=\la f|1\ra S_{11}\la 1|i\ra, \quad 
\Phi^{f\leftarrow i}_{22}=\la f|2\ra S_{22}\la 2|i\ra. 
\end{eqnarray}
The other two paths are those in which the system jumps between the states,
and correspond to $\Delta n=-1$ and $\Delta n=1$, respectively.
Their probability amplitudes are
\begin{eqnarray} \label{3.7}
\Phi^{f\leftarrow i}_{12}=\la f|1\ra S_{12}\la 1|i\ra, \quad
\Phi^{f\leftarrow i}_{21}=\la f|2\ra S_{21}\la 2|i\ra. 
\end{eqnarray}
For the probabilities of the three outcomes we, therefore, have
\begin{eqnarray} \label{3.8}
P^{f\leftarrow i}(-1)={|\Phi^{f\leftarrow i}_{12}|^2}/[{|\Phi^{f\leftarrow i}_{11}+\Phi^{f\leftarrow i}_{22}|^2+|\Phi^{f\leftarrow i}_{12}|^2+|\Phi^{f\leftarrow i}_{21}|^2}]
\\ \nonumber
P^{f\leftarrow i}(0)={|\Phi^{f\leftarrow i}_{11}+\Phi^{f\leftarrow i}_{22}|^2}/[{|\Phi^{f\leftarrow i}_{11}+\Phi^{f\leftarrow i}_{22}|^2+|\Phi^{f\leftarrow i}_{12}|^2+|\Phi^{f\leftarrow i}_{21}|^2}]
\\ \nonumber
P^{f\leftarrow i}(1)={|\Phi^{f\leftarrow i}_{21}|^2}/[{|\Phi^{f\leftarrow i}_{11}+\Phi^{f\leftarrow i}_{22}|^2+|\Phi^{f\leftarrow i}_{12}|^2+|\Phi^{f\leftarrow i}_{21}|^2}],
\end{eqnarray}
which sum, as they should, to unity.

\section {The three-box case}
Next we apply  the path analysis to the original three-box
'paradox' as presented in \cite{3B1,ABOOK,3B2}.
A three-state system with a zero Hamiltonian,
$\hat{H}\equiv 0$ in prepared  at $t=0$ in the state 
\begin{equation} \label{3.1a}
|i\ra =3^{-1/2}(|1\ra + |2\ra + |3\ra)
\end{equation}
and then at $t=T$ found in the state
\begin{equation} \label{3.2a}
|f\ra =3^{-1/2}(|1\ra + |2\ra - |3\ra)
\end{equation}
where the eigenstates $|n\ra$, $n=1,2,3$ of the 'position' operator $\hat{n}=\sum_{n=1}^3 |n\ra n \la n| $
 form an orthonormal basis.
Following  \cite{3B1,ABOOK,3B2}, at some intermediate time, say, $t=T/2$ we will wish to perform  an impulsive von Neumann 
measurement of two commuting operator functions of $\hat{n}$ (the last expression is the matrix form in the chosen basis)
\begin{equation} \label{3.3a}
\hat{P}_1\equiv  |1\ra  \la 1| = diag(1,0,0),
\end{equation}
\begin{equation} \label{3.4a}
\hat{P}_2\equiv  |2\ra  \la 2| = diag(0,1,0).
\end{equation}
>From Eq.(\ref{0.2}) we note that since $\hat{H}\equiv 0$ there are just 
three constant virtual paths 
$ n(t)\equiv 1, \quad 2, \quad 3$
with the probability amplitudes
\begin{eqnarray} \label{3.7a}
\Phi^{f \leftarrow i}_{1}=\la f|1\ra \la 1|i\ra = 1/3\\
\nonumber
\Phi^{f \leftarrow i}_{2}=\la f|2\ra \la 2|i\ra = 1/3\\
\nonumber
\Phi^{f \leftarrow i}_{3}=\la f|3\ra \la 3|i\ra = -1/3
\end{eqnarray}
which connect  the initial and final 
states (\ref{3.1a}) and (\ref{3.2a}) .
Now the projector $\hat{P}_1$ in Eq.(\ref{3.3a}) is a function of $\hat{n}$ with and
eigenvalue $F(1)=1$ and another doubly degenerate eigenvalue $F(2)=F(3)=0$.
Thus, as discussed in Sect.2, its measurement creates two real pathways between $|i\ra$ and $|f\ra$, $\{1\}$ and $\{2+3\}$ and using Feynman's rule to assign  probabilities
we write
\begin{eqnarray} \label{3.8a}
P^{f \leftarrow i}_{\{1\}}=
|\Phi^{f \leftarrow i}_1|^2/
\{|\Phi^{f \leftarrow i}_{1}|^2+|\Phi^{f \leftarrow i}_{2}+\Phi^{f \leftarrow i}_{3}|^{2}\}
=1\\
P^{f \leftarrow i}_{\{2+3\}}=
 |\Phi^{f \leftarrow i}_{2}+\Phi^{f \leftarrow i}_{3}|^{2}/
\{|\Phi^{f \leftarrow i}_{1}|^2+|\Phi^{f \leftarrow i}_{2}+\Phi^{f \leftarrow i}_{3}|^{2}\}
=0 
\end{eqnarray}
so that the route combining the virtual paths $2$ and $3$ is never travelled
due to the special choice of the initial and final states (\ref{3.1}) and (\ref{3.2}) (Fig.1b).  
Similarly, a measurement of the projector onto the second state, $\hat{P}_2$, creates two real pathways,
$\{2\}$ which is travelled with certainty and $\{1+3\}$ which is never travelled (Fig.1c)..
A simple look at the probabilities to arrive 
 in other final states reveals that the two measurements
 convert,  as was discussed in Sect.3, the original unobserved system into
 two essentially different classical networks.
 Consider, therefore, two other members of the orthonormal set $\{|f\ra, |g\ra, |h\ra\}$,
 which can be chosen, for example, as
 \begin{equation} \label{3.8b}
|g\ra =2^{-1/2}(|2\ra +|3\ra)
\end{equation}
\begin{equation} \label{3.9}
|h\ra =6^{-1/2}(-2|1\ra + |2\ra - |3\ra).
\end{equation}
The probabilities
given in the Table 1 show that the only property the two networks shown in Fig.1b and Fig.1c share is 
the same probability to reach
$|f\ra$ from $|i\ra$, while for the final states $|g\ra$ and $|h\ra$ the probabilities differ. 
Thus the above results are no more 'paradoxical' than the fact that 
two otherwise different classical stochastic networks may have the same probability
associated with two different pathways.
\newline 
Finally, much attention has been paid \cite{ABOOK} to the failure of the product rule,
i.e., the fact that in the three-box case measurement of either $\hat{P}_1$ or 
$\hat{P}_2$ always yields the value $1$, a measurement of their product,
\begin{equation} \label{3.5a}
\hat{P}_1\times \hat{P}_2 = diag(0,0,0),
\end{equation}
is certain to yield $0$ and not $1$. There are two aspects to this question.
Firstly, for the {\it same} statistical ensemble the product of two sharply defined quantities  with
values, say, $a$ and $b$ is also sharply defined with the value $ab$.
There is no contradiction, since the failure of the product rule only serves to illustrate further the fact that 
a new statistical ensemble is created with each choice of measured quantities
even if operators for all quantities commute.
In particular, since all eigenvalues of $\hat{P}_1\times \hat{P}_2$
are the same, no new real pathways are created by its measurement and the corresponding 
statistical ensemble is that of a system without intermediate measurements,
essentially different from the ensembles produced by a measurement of $\hat{P}_1$ or 
$\hat{P}_2$.  
Secondly, the product rule is recovered if a system 
is prepared in an initial state $|i\ra$ but no post-selection is performed \cite{ABOOK},
 and it is instructive to see how this happens.
As the three box example shows,
for a transition into a single final state, we have the freedom of choosing 
both $|i\ra$ and $|f\ra$ in such a way that $\hat{P}_1$, $ \hat{P}_2$ and $\hat{P}_1\times \hat{P}_2$ are sharply defined, yet the product rule does not apply.
The pre-selected only average of an operator $F(\hat{n})$ is obtained by averaging
over all real pathways leading to a given final state and then over all final states,
\begin{equation} \label{3.6a}
\la F(\hat{n}) \ra \equiv \sum_z \sum_{m=1}^M F_m |\la z|\hat{P}_m|i\ra|^2 = \sum_{n=1}^N
F(n)|\la n|i\ra|^2.
\end{equation}
Thus for the variable $F$ to be sharply defined in a pre-selected only ensemble, 
it must be sharply defined for the transitions into all final states
and not just into one chosen $|f\ra$. This is a far more restrictive
requirement leaving only the freedom of choosing $|i\ra$,  which
is  not enough to violate the product rule. Indeed, in the three box case both $\hat{P}_1$ and  $\hat{P}_2$ are sharply defined only for $|i\ra=|1\ra$, $\la \hat{P}_1\ra=1$, $\la \hat{P}_2\ra=0$;
$|i\ra=|2\ra$,  $\la \hat{P}_1\ra=0$, $\la \hat{P}_2\ra=1$ and $|i\ra=|3\ra$,  $\la \hat{P}_1\ra=0$, $\la \hat{P}_2\ra=0$. In all three cases the product $\hat{P}_1\times \hat{P}_2$ and has a value
$0$ which equals the product of the values of $\hat{P}_1$ and $\hat{P}_2$.

\section {A classical model. Invasive observations and non-locality}
The authors of Ref.\cite{3B2} described the three-box case as a quantum paradox 
in the 'sense that it is a classical task which cannot be accomplished using classical means'
and one wonders what, if anything, would constitute its closest classical analogue. For such an analogue we consider a classical system where a ball can reach three final destinations
$f$, $g$ and $h$ by rolling down the system of tubes shown in Fig.\ref{CLASS}.
With equal probabilities $p_0=1/3$ the ball reaches one of its intermediate positions (boxes)
$1$, $2$ and $3$ and then is redirected to the final states with the branching probabilities
$p(k,z)$, $k=1,2,3$, $z=f,g,h$. The properties of the network are completely determined by
the probabilities $p^{z\leftarrow i}_{\[k\]}=p_0p(k,z)$ for travelling the nine pathways available to the
ball. In particular, probability for an observable property (e.g., arrival in the final state $f$) is found by summing
the probabilities for all pathways that share the property. It is a simple matter to see that the
existence of such classical
probabilities unchanged by an observation
is incompatible with the probabilities prescribed for the three-box case by the Feynman' rule (II). 
The check consists in equating the classical result with the quantum one and eventually arriving
at a contradiction. In particular, we have
\begin{eqnarray} \label{A1}
p^{f\leftarrow i} \equiv  p^{f\leftarrow i}_{\[1\]}+ p^{f\leftarrow i}_{\[2\]}+ p^{f\leftarrow i}_{\[3\]}=
|\la f|i\ra|^2 = 1/9\\ 
\nonumber
\quad (no\q box\q is \q opened)\q 
\end{eqnarray}
\begin{eqnarray} \label{A2}
p^{f\leftarrow i}_{\[1\]}=|\la f|1\ra\la1|i\ra|^2 = 1/9\\
\nonumber
 \quad (box \q1\q opened;\q ball\q found)\q
 \end{eqnarray}
 \begin{eqnarray} \label{A3}
p^{f\leftarrow i}_{\[2\]}=|\la f|2\ra\la2|i\ra|^2 = 1/9\\ 
\nonumber
\quad (box\q 2\q opened;\q ball\q found)\q
\end{eqnarray}
\begin{eqnarray} \label{A4}
p^{f\leftarrow i}_{\[2+3\]}\equiv p^{f\leftarrow i}_{\[2\]}+p^{f\leftarrow i}_{\[3\]}=
|\la f|2\ra\la2|i\ra+\la f|3\ra\la3|i\ra|^2=0\\
\nonumber
 \quad (box\q 1\q opened;\q ball \q not\q found)\q 
  \end{eqnarray}
 \begin{eqnarray} \label{A5}
p^{f\leftarrow i}_{\[1+3\]}\equiv p^{f\leftarrow i}_{\[1\]}+p^{f\leftarrow i}_{\[3\]}=
|\la f|1\ra\la1|i\ra+\la f|3\ra\la3|i\ra|^2=0\\
\nonumber
 \quad (box\q 2\q opened; \q ball \q not \q found).\q
\end{eqnarray}
Clearly equations (\ref{A1}-\ref{A5}) cannot be solved for $0\le p^{i\leftarrow i}_{[k]}\le 1$.
 Indeed, while  Eq.(\ref{A2}) sets the value  
$p^{f\leftarrow i}_{\[1\]}=1/9$, Eq.(\ref{A5}) implies $p^{f\leftarrow i}_{\[1\]}=0$
which must, therefore, be changed.
This demonstrates that (just as an electron in the double slit experiment) the quantum
particle not found in the first box does not travel either of the two remaining pathways
with any particular probability. Thus, to make the system in Fig.\ref{CLASS} mimic
the quantum results we must make an observation invasive, i.e., ensure that the mere act of opening a box would change
the statistical ensemble. For example, we may equip each box with a photo-element,
illuminated when the box is opened,  which we would then use to send a signal to alter the branching probabilities $p(k,z)$. We note further that such an interaction must be non-local in the sense that
$p(k,z)$ must  be changed not just for the box we open, but even for the one
we never look into. To prove this point, assume for a moment that the probabilities
are altered only for the box whose lid is opened and are $\tilde{p}^{f\leftarrow i}_{\{k\}}$ with the 
lid down and $p^{f\leftarrow i}_{\{k\}}$ with the lid up, respectively. Then from  Eq.(\ref{A1}) we have
\begin{equation} \label{A6}
  \tilde{p}^{f\leftarrow i}_{\[1\]}+ \tilde{p}^{f\leftarrow i}_{\[2\]}+  \tilde{p}^{f\leftarrow i}_{\[3\]}=1/9
\end{equation}
in a clear contradiction to Eqs. (\ref{A4}) and (\ref{A5}) which together imply that 
\begin{equation} \label{A7}
  \tilde{p}^{f\leftarrow i}_{\[1\]}+ \tilde{p}^{f\leftarrow i}_{\[2\]}+ \tilde{p}^{f\leftarrow i}_{\[3\]}=
 0.
\end{equation}
Table \ref{T2} gives possible values of observation-dependent  branching probabilities with which the statistical properties of the network in Fig.\ref{CLASS}
are equivalent to those of a quantum three-box system for which either of the three boxes 
can be opened.

\section{Conclusions and discussion}
In summary, we argue that
whenever a set of measurements is performed on a quantum system,
the experimentalist studies the properties of a classical statistical
ensemble defined by conditional probabilities for all possible outcomes.
Such an ensemble can be represented as a network of real pathways
connecting the initial and final states, with the probability to travel a particular
path given by the Feynman's rule quoted in the Introduction.
The principal difference with the classical case is that since quantum
meter may perturb the measured system, each choice of the 
measured quantities fabricates its own ensemble
in such a way that the properties of one ensemble cannot be attributed to another ensemble
or to the original unobserved system.  
The three-box example provides a simple illustration of the above:
a decision to open the second instead of the first box alters the probabilities
to arrive into final sates $|g\ra$  and $|h\ra$, something that would never happen with non-invasive
observations performed on the same statistical ensemble.
Similarly, the failure of the product rule for the measurements of the 
projectors $\hat{P}_1$,  $\hat{P}_2$ and of their product  also indicates that
the three measurements correspond to three different ensembles.
\newline The discussion of whether the three box case deserves the status of a  "paradox" is not yet over. 
The statements in favour of such a status include " the peculiarity of having the particle in several places simultaneously in a stronger sense than it is in a double slit experiment" \cite{ABOOK} and
'a paradox ... is that at a particular time... a particle is in some sense both with certainty in one box, and with certainty in another box'\cite{3B2}. To clarify the sense in which that happens,
one can imagine two classical model of Sect.5 with the branching probabilities set to the values
given in the first and the second rows of Table 2, respectively. Post-selecting the ball in the final
state $f$, one observes that the ball passes with certainty through box 1 {\it if} it was put into the  first model, and with certainty in box 2 {\it if} the second model is chosen. Precisely the same
can be said about the particle in the quantum three-box case: whether the two classical 
networks were prefabricated, or whether one of the two is fabricated while the ball is 
in motion is of no importance. In this sense our analysis is  closest to that of  Kastner \cite{Kast}, who criticised the counterfactual use of the ABL rule by pointing out that conflicting properties "cannot apply to 
the same individual particle" and  to a more general study of both pre- and post-selected quantum ensembles conducted by Bub and Brown in Ref.\cite{BB}. 
\newline
The authors of \cite{3B2} also characterised the three-box case as "a classical task which cannot be be accomplished using classical means".  The implication seems to be that the task cannot 
be accomplished by a passive observer who is not allowed to change the statistical properties 
of the system in the course of an observation, i.e., denied the right to do
precisely what is done by a quantum meter. We find a different line of thought more natural: 
comparing probabilities for the classical model in Fig.\ref{CLASS} with those prescribed
by quantum mechanics for the three-box case one would conclude that something 
must alter the properties of the system while it is observed and that effect of such a perturbation 
cannot be limited to just the box that has been opened.
A quick consultation with a quantum mechanical text book would, indeed,  confirm that quantum measurements are by nature invasive and that quantum correlations are inherently non-local
(see, for example Ref.\cite{GRIB}). We leave it to the reader to decide whether the situation merits the title of a paradox, in particular in view of Finkelstein's  observation \cite {Fink}
that the notion of a paradox is largely 'a matter of personal psychology'.
\newline
Finally we note that Feynman's approach with its notion of interfering
scenarios being converted into exclusive ones captures the main features of the 
three box "paradox" and helps simplify the language of the discussion.  
Although the use of Feynman path integral does not, in itself, add further substance to the discussion, 
it is extremely useful for book-keeping purposes.
A complete set of Feynman paths (in the three-box case nine, counting all final states) represents elementary histories which an intermediate measurement combines into a number of real pathways. 
Once the network of real pathways are identified, the Feynman recipe (II) provides the 
frequencies with which each pathway is travelled. This is the only place where quantum
mechanics enters into what subsequently becomes a simple exercise in purely classical probabilities. 
Any conditional probability can now be found by comparing the frequences  which gives, as it should, 
the same result as applying the Bayes formula Refs.\cite{3B1}, \cite{Leif}, \cite {Kirk2},
 but without ambiguities which may arise in such an application \cite{Kirk4}.  
Thus, the task of assigning probabilities to a pre- and post-selected quantum system
 reduces to applying a minimalist set of Feynman's rules  which, in turn, helps avoid over-interpretation 
which might occur if more esoteric formulations of quantum 
mechanics \cite{3B1}, \cite{ABOOK}  are employed.

\section{Acknowledgements}

R. Sala Mayato is grateful to acknowledge Ministerio de Educaci\'on y Ciencia,
Plan Nacional under grants No. FIS2004-05687 and No. FIS2007-64018, 
and Consejer\'\i a de Educaci\'on, Cultura
y Deporte, Gobierno de Canarias under grant PI2004-025. 
I. Puerto is greatful to acknowledge Ministerio de Educaci\'on y Ciencia under grant
AP-2004-0143.
D. Sokolovski is greatful to acknowledge Universidad de La Laguna for partial support under
project ``Ayudas para la estancia de profesores e investigadores no vinculados a la Universidad de La Laguna''.

\newpage

\begin{figure}
\vskip0.5cm
\includegraphics[width=5.cm]{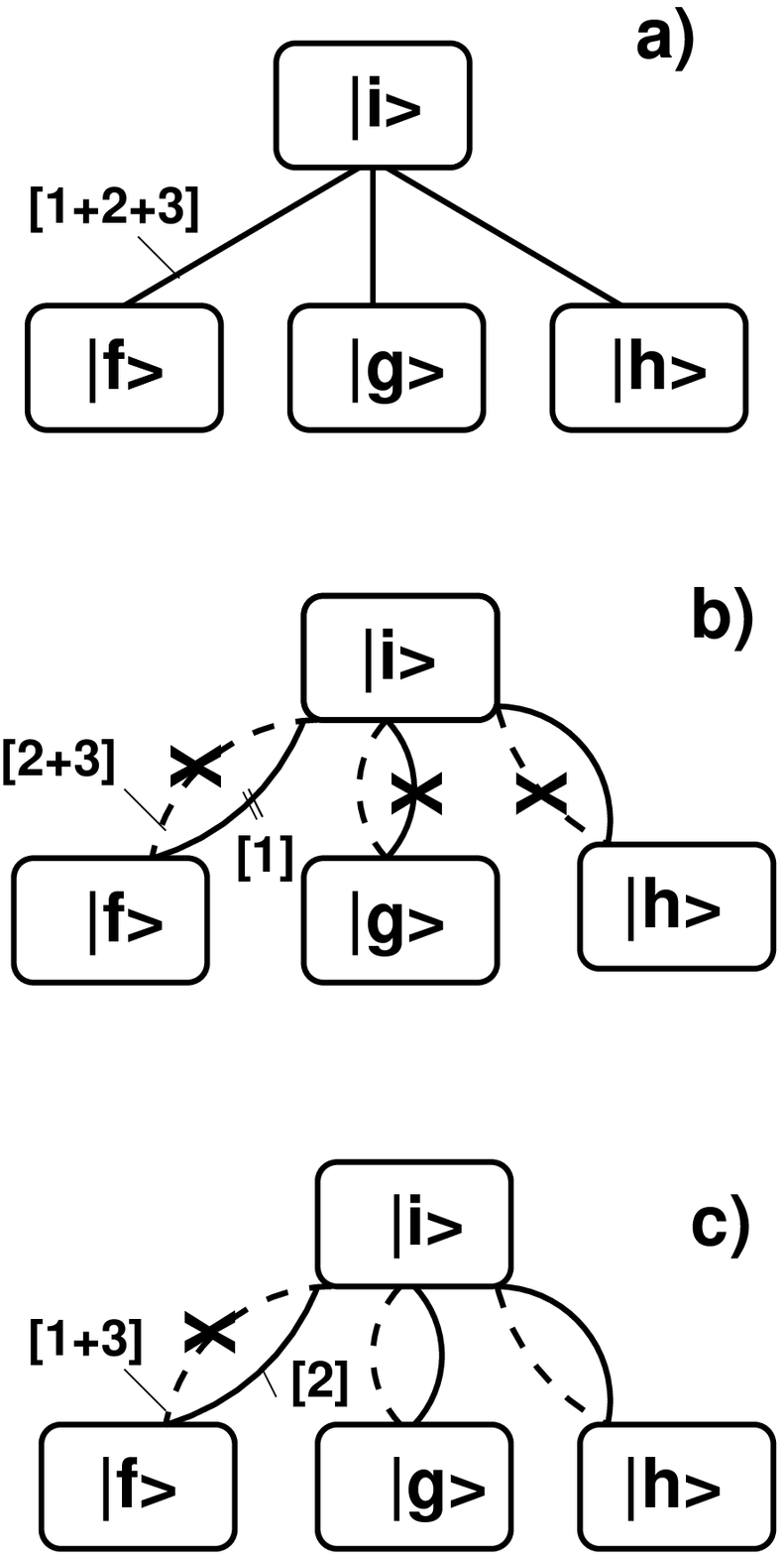}
\caption {Real pathways produced by intermediate measurements on a three-state system.
\newline
a) Nothing or  $\hat{P}_1\hat{P}_2$ is measured;
b)  $\hat{P}_1$ is measured;
c) $\hat {P}_2$ is measured.
\newline A cross indicates that a route is not travelled.}
\label{f1}
\end{figure}
\begin{figure}
\vskip0.5cm
\includegraphics[width=7cm]{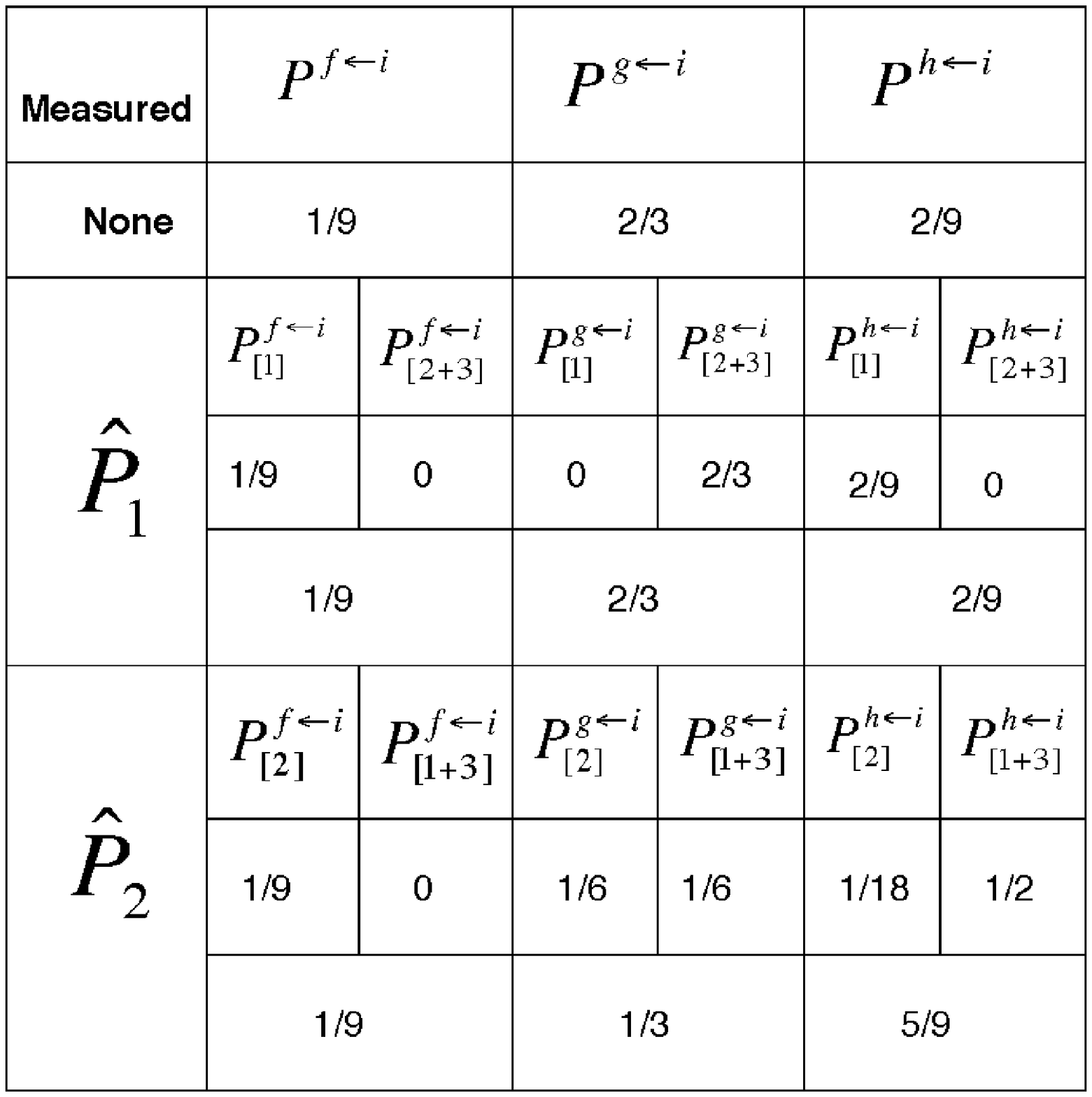}
\caption {Probabilities for real pathways shown in Fig.\ref{f1}}.
\label{TABLE}
\end{figure}

\newpage
\begin{figure}
\vskip0.5cm
\includegraphics[width=8cm]{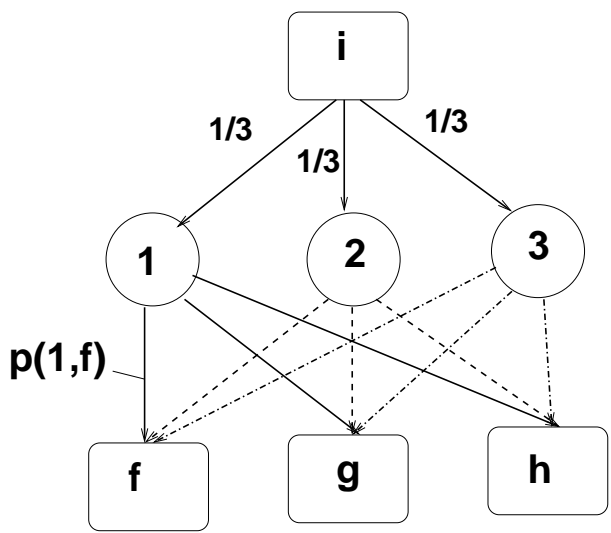}
\caption {A classical three-box model.}
\label{CLASS}
\end{figure}

\begin{figure}
\vskip0.5cm
\includegraphics[width=15cm]{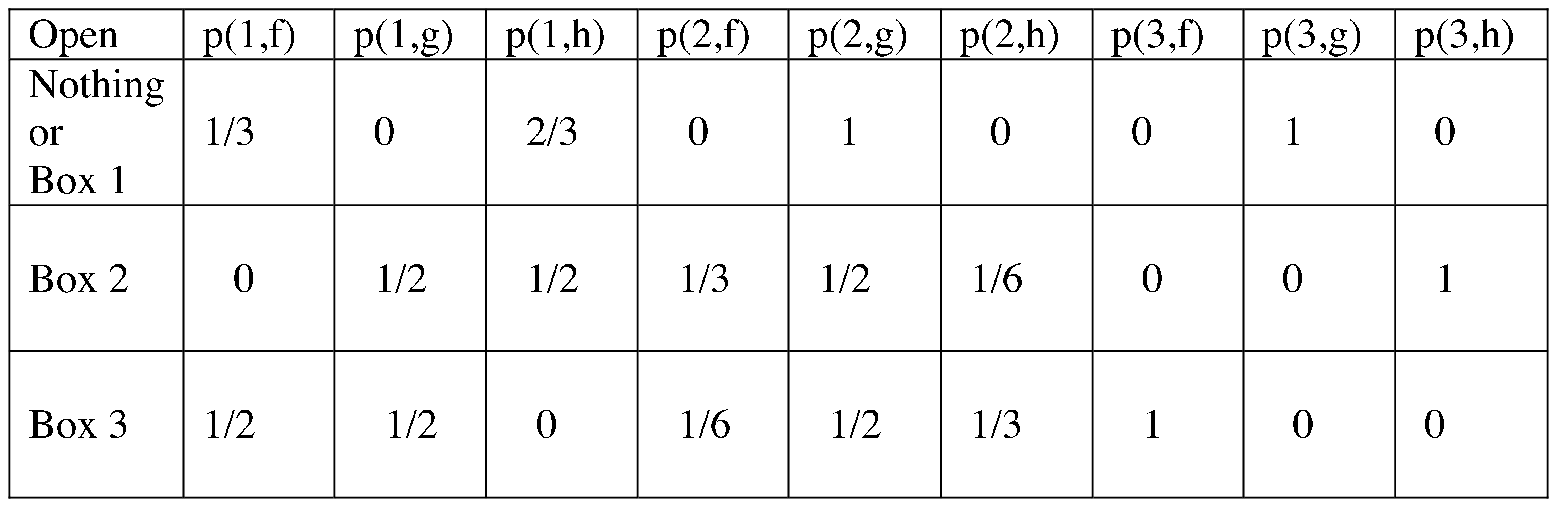}
\caption {Table 2.  Branching probabilities for the model in Fig.\ref{CLASS} .}
\label{T2}
\end{figure}

\end{document}